\documentclass[twocolumn,english,showpacs,preprintnumbers,amsmath,amssymb]{revtex4}
\usepackage[T1]{fontenc}
\usepackage[latin9]{inputenc}
\usepackage{graphicx}
\usepackage{esint}

\makeatletter
\@ifundefined{textcolor}{}
{%
 \definecolor{BLACK}{gray}{0}
 \definecolor{WHITE}{gray}{1}
 \definecolor{RED}{rgb}{1,0,0}
 \definecolor{GREEN}{rgb}{0,1,0}
 \definecolor{BLUE}{rgb}{0,0,1}
 \definecolor{CYAN}{cmyk}{1,0,0,0}
 \definecolor{MAGENTA}{cmyk}{0,1,0,0}
 \definecolor{YELLOW}{cmyk}{0,0,1,0}
 }

\makeatother

\usepackage{babel}

\begin{document}

\title{Supersolid Phase of Cold Fermionic Polar Molecules in 2D Optical
Lattices}

\author{Liang He and Walter Hofstetter}

\affiliation{Institut f\"{u}r Theoretische Physik, Johann Wolfgang Goethe\textendash{}Universit\"{a}t,
60438 Frankfurt/Main, Germany }
\begin{abstract}
We study a system of ultra-cold fermionic polar molecules in a two-dimensional
square lattice interacting via both the long-ranged dipole-dipole
interaction and a short-ranged on-site attractive interaction. Singlet
superfluid, charge density wave, and supersolid phases are found to
exist in the system. We map out the zero temperature phase diagram
and find that the supersolid phase is considerably stabilized by the
dipole-dipole interaction and thus can exist over a large region of
filling factors. We study the melting of the supersolid phase with
increasing temperature, map out a finite temperature phase diagram
of the system at fixed filling, and determine the parameter region
where the supersolid phase can possibly be observed in experiments.
\end{abstract}


\maketitle

\section{Introduction}

Ultracold atom physics has undergone spectacular development in the
last two decades. During the last years, after the first theoretical
proposal to simulate the Bose-Hubbard model with ultracold bosons
in an optical lattice \cite{BH_The} and its successful experimental
realization \cite{BH_Exp}, considerable theoretical and experimental
progress has been made on degenerate quantum gases in optical lattices
\cite{Lattice_review}. One major common goal shared by these efforts
is to simulate typical systems or minimal models of condensed matter
physics, e.g. the Hubbard model, using ultracold atoms confined in
optical lattices \cite{Fermi_Hubbard_The,Fermi_Hubbard_Exp}. It is
expected that these investigations can give key insights into unresolved
open questions in condensed matter physics, e.g. the mechanism of
high-$T_{c}$ superconductivity. Usually, interactions in cold gases
are isotropic and short ranged. Naturally one can ask the question
which type of new physics would arise in these systems if the atoms
have additional long-range anisotropic interactions. 

As a matter of fact, a wealth of interesting physics has already been
revealed by theoretical and experimental investigations of ultracold
dipolar Bose gases both in harmonic traps \cite{dip_the_1,dip_the_2,dip_the_3,dip_the_4,dip_the_5,dip_exp1,dip_exp2,dip_exp_3,dip_exp_4}
and optical lattices \cite{dip_bose_lat_mf_1,dip_bose_lat_mf_2,dip_bose_square,dip_bose_tri}.
In contrast to dipolar Bose gases, due to the Pauli exclusion principle,
dipolar effects in fermionic systems can manifest themselves only
if the dipolar interaction energy is at least of the order of the
Fermi energy. Although there have been already some theoretical investigations
on these systems \cite{dip_fermi_1,dip_fermi_2,dip_fermi_3,dip_fermi_4,dip_fermi_5},
this condition could hardly be reached experimentally until the recent
successful realization of a degenerate quantum gas of fermionic $^{40}$K$^{87}$Rb
polar molecules \cite{PM_Exp}. With this achievement, the door towards
exploring the many-body physics originating from dipole-dipole interactions
in fermionic systems has been opened, which has already lead to a
number of theoretical investigations on these systems \cite{dip_fer_molecule_1,dip_fer_molecule_2,dip_fer_molecule_3,dip_fer_molecule_4,dip_fer_molecule_5,dip_fer_molecule_6,dip_fer_molecule_7,dip_fer_molecule_8,dip_fer_molecule_9,dip_fer_molecule_10,dip_fer_molecule_11,dip_fer_molecule_12,dip_fer_molecule_13,dip_fer_molecule_14,dip_fer_molecule_15,dip_fer_molecule_16,dip_fer_molecule_17,dip_fer_molecule_18,dip_fer_molecule_19}.
Motivated by experimental progress on ultracold polar molecules, here
we investigate the low temperature quantum phases of a system of cold
polar fermions loaded into a 2D square optical lattice. 

The rest part of this paper is organized as follows. In Sec. II, we
introduce the system and the model studied here. In Sec. III, we describe
the theoretical approach used in our investigation. In Sec. IV, the
main section of this paper, we present a detailed discussion of the
quantum phases at both zero and finite temperature. Finally, we conclude
in Sec. V.

\section{System and Model}

Motivated by ongoing experiments, we focus our investigations on possible
setups based on $^{40}$K$^{87}$Rb molecules which have a permanent
electric dipole moment of $0.57$ Debye (D), where $1\mathrm{D}=3.336\times10^{-30}\mathrm{C\cdot m}$.
However the effective dipole moment in the laboratory frame is zero
in the absence of an external electric field. When an external electric
field is applied the molecules align with the field and have an induced
dipole moment $d$ which increases with the strength of the external
field. Currently, the experimentally accessible range of $d$ is $[0,0.22\, D]$
\cite{mol_dip_coll}. In this work we assume the external electric
field is oriented perpendicular to the optical lattice plane. 

Loading the molecules into an optical lattice, the physics of this
system for sufficiently low filling can be captured by an extended
Hubbard model within the lowest band approximation,\begin{eqnarray}
H & = & -t\sum_{<i,j>,\sigma}c_{i\sigma}^{\dagger}c_{j\sigma}+U\sum_{i}n_{i\uparrow}n_{i\downarrow}\label{eq:Hami}\\
 &  & +\frac{c_{d}}{2}\sum_{i\neq j;\sigma\sigma'}\frac{n_{i\sigma}n_{j\sigma'}}{|\mathbf{R}_{i}-\mathbf{R}_{j}|^{3}}-\mu\sum_{i\sigma}n_{i\sigma}\,.\nonumber \end{eqnarray}
Here, $\langle i,j\rangle$ denotes nearest-neighbour sites, $c_{i\sigma}^{\dagger}$($c_{i\sigma}$)
is the creation (annihilation) operator of a fermionic molecule with
spin $\sigma$ on site $i$ in the Wannier representation, and $n_{i\sigma}=c_{i\sigma}^{\dagger}c_{i\sigma}$
is the particle number operator. The first term in Eq. ($\ref{eq:Hami}$)
describes the kinetic energy with $t$ being the hopping amplitude;
the second term represents the onsite interaction $U$ between molecules
with opposite spin; the third term originates from the long-range
dipole-dipole interactions, where $c_{d}=d^{2}/(4\pi\varepsilon_{0}a^{3})$
characterizes the dipole-dipole interaction strength with $a$ and
$\varepsilon_{0}$ being the lattice constant and the vacuum permittivity
respectively, and $\mathbf{R}_{i}$ is the dimensionless position
vector of lattice site $i$; finally, in the last term $\mu$ denotes
the chemical potential. 

In our calculations we assume a 2D optical lattice created by laser
beams with wavelength $\lambda=1064\mathrm{nm}$ as used in the experimental
setup of Ref. \cite{mol_dip_coll}. We consider the following parameters:
the height of the lattice potential in the direction perpendicular
to the lattice plane $V_{z}^{\mathrm{lat}}$ and parallel to the plane
$V_{\perp}^{\mathrm{lat}}$ are chosen as $V_{\perp}^{\mathrm{lat}}=10E_{R}$,
$V_{z}^{\mathrm{lat}}=10V_{\perp}^{\mathrm{lat}}$, where $E_{R}=h^{2}/(2m\lambda^{2})$
is the recoil energy, $m$ is the mass of molecule and $h$ is Planck's
constant. Under these conditions, the dipolar interaction strength
is in the range $[0,\,0.03E_{R}]$, and the hopping amplitude is approximately
$0.2E_{R}$. Therefore the ratio between the dipolar interaction strength
and the hopping amplitude $c_{d}/t$ is approximately in the range
$[0,0.15]$, which we assume in our calculations.

A good estimate of the strength of the onsite interaction $U$ requires
a detailed study of all types of short-range interactions between
the polar molecules, which is beyond the scope of this work. Since
the main aim of this work is to investigate the possibility of observing
a supersolid phase of polar molecules, we assume that the onsite interaction
is attractive, i.e., $U<0$. Moreover, we notice that since large
attractive on-site interactions will make the system unstable, the
region of $|U|$ investigated in this work is restricted to $[t,\,8t]$. 

On the route towards observing the supersolid phase in experiments,
the temperature plays a dominant role. In current experiments with
polar molecules in a harmonic trap, the lowest temperature which can
be reached is of the order of the Fermi temperature $T_{F}$ \cite{mol_dip_coll}.
Also, when the lattice potential is ramped up, generally, the molecules
will be heated in this process, which results in even higher temperatures.
In the following, the system is studied at both zero and finite temperatures,
and the temperature region where a supersolid phase can be possibly
observed is determined.

\section{Method}

Before discussing our results in detail, we give a brief description
of our theoretical approach. In polar molecule systems, we have to
consider not only the on-site attraction of the molecules but also
the inter-site repulsion due to the dipolar interaction. For treating
the on-site interactions, the Dynamical Mean-Field Theory (DMFT) \cite{DMFT}
is well suited since it is non-perturbative, captures the local quantum
fluctuation exactly, and gives exact results in the infinite-dimensional
limit. For long-ranged interactions in fermionic system, on the other
hand, there are, to our knowledge, up to now only few efficient ways
to treat them, although DMFT has already been generalized to the so
called cellular DMFT in order to treat short-range inter-site interactions
\cite{CDMFT}.

We include long-range interactions as follows. First we notice that
in the high-dimensional limit inter-site interactions only contribute
on the Hartree level \cite{LD_HFA}. In other words, the Hartree term
of the inter-site interaction will dominate as the spatial dimension
of the system increases. This motivates us to keep only the Hartree
contribution of the inter-site interaction in the Hamiltonian as an
approximation to the original Hamiltonian (\ref{eq:Hami}), i.e. \begin{eqnarray}
 &  & \frac{c_{d}}{2}\sum_{i\neq j;\sigma\sigma'}\frac{n_{i\sigma}n_{j\sigma'}}{|\mathbf{R}_{i}-\mathbf{R}_{j}|^{3}}\\
 & \simeq & \sum_{i\neq j,\sigma\sigma'}c_{d}\frac{1}{|\mathbf{R}_{i}-\mathbf{R}_{j}|^{3}}\langle n_{j\sigma}\rangle(n_{i\sigma'}-\frac{1}{2}\langle n_{i\sigma'}\rangle).\label{eq:HF_deouple}\end{eqnarray}
Our DMFT Hamiltonian therefore takes the form \begin{equation}
H =-t\sum_{\langle i,j\rangle,\sigma}c_{i\sigma}^{\dagger}c_{j\sigma}+U\sum_{i}n_{i\uparrow}n_{i\downarrow}+\sum_{i\sigma}(\tilde{V}_{i}-\mu)n_{i\sigma}
\label{eq:HF_Hamiltonian}\end{equation}
where\begin{equation}
\tilde{V}_{i}=\sum_{j\,(i\neq j),\sigma}c_{d}\frac{1}{|\mathbf{R}_{i}-\mathbf{R}_{j}|^{3}}\langle n_{j\sigma}\rangle\label{eq:V_hf}\end{equation}
and we have rescaled the chemical potential $\mu$ in (\ref{eq:HF_Hamiltonian}) according to the trivial constant term in (\ref{eq:HF_deouple}).

We investigate this system using real-space DMFT (R-DMFT)\cite{R-DMFT_1,R-DMFT_2},
which is an extension of DMFT to a position-dependent self-energy
and fully captures the inhomogeneity of the system. Within DMFT/R-DMFT,
the physics on each lattice site is determined from a local effective
action which can be captured by an effective Anderson impurity model
\cite{DMFT}. In this work, we use Exact Diagonalization (ED) \cite{ED_solver_1,ED_solver_2}
of the Anderson impurity Hamiltonian to solve the local action. Details
of the R-DMFT method can be found in previous works \cite{R-DMFT_1,R-DMFT_2}. 

Within each R-DMFT iteration, the Hartree contributions $\tilde{V}_{i}$
are calculated and the Hamiltonian (\ref{eq:HF_Hamiltonian}) is updated
according to the new values $\tilde{V}_{i}$. This modified R-DMFT
iteration is repeated until a convergent solution is obtained.

\section{RESULTS}

\subsection{Zero temperature }

\subsubsection{Phase diagram}

\begin{figure}
\includegraphics[clip,width=3in]{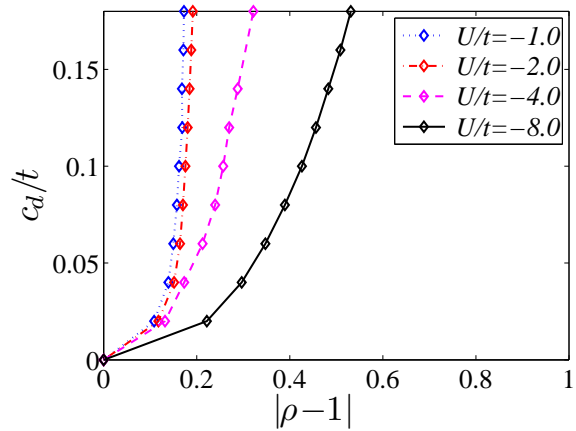}

\caption{(Color online). Phase diagram of the system with respect to the filling
per lattice site $\rho$ and the strength of dipolar interaction $c_{d}$,
obtained from calculations on a $12\times12$ square lattice. Different
curves indicate the phase boundaries between supersolid (SS) and homogeneous
singlet superfluid (HSF) at different onsite interaction strengths
$U$. On the left-hand side of each curve the system is in the SS
phase (except exactly at half filling ($\rho=1.0$)) while the right-hand
side region indicates the homogeneous singlet superconducting phase.
From left to right, different curves correspond to $U/t=-1.0,-2.0,-4.0$,
and $-8.0$ respectively.}

\label{Flo:pd_12x12}
\end{figure}

At zero temperature, the main properties of the system are summarized
in the phase diagram in Fig.~\ref{Flo:pd_12x12}, which shows the
phase boundary between the supersolid (SS) phase and the homogeneous
singlet superfluid phase (HSF) at different onsite interaction strengths
$U$ in terms of the filling factor $\rho\equiv\sum_{i\sigma}\langle n_{i\sigma}\rangle/N$
and the dipolar interaction strength $c_{d}$, where $N$ is the number
of lattice sites. This phase diagram is based on calculations for
a $12\times12$ square lattice. We will denote the phase boundary
between the supersolid (SS) phase and the homogeneous singlet superfluid
phase the \textit{SS-HSF boundary} in the following. 

Along the vertical line at half filling ($\rho=1$) in Fig.~\ref{Flo:pd_12x12},
irrespective of the strengths of on-site and inter-site interactions,
the system is always in an incompressible charge density wave (CDW)
phase, which is characterized both by vanishing compressibility $\partial\rho/\partial\mu=0$
and a finite density modulation in real space. More specifically,
the density distribution over the lattice has a checkerboard (CB)
structure, which is characterized by $\rho_{\mathbf{Q}}\equiv$$\sum_{j}e^{i\mathbf{Q}\cdot\mathbf{R}_{j}}\rho_{j}/N$
with $\rho_{j}\equiv\sum_{\sigma}\langle n_{j\sigma}\rangle$ and
$\mathbf{Q}=(\pi,\pi)$. In our calculations, we choose $\rho_{\mathbf{Q}}$
as the CDW order parameter and we will denote this incompressible
CDW phase as {}``CB solid''.

Away from half filling but still in the region bounded by the SS-HSF
boundary and the vertical line of half filling, a supersolid phase
is obtained by doping the CB solid with either particles or vacancies,
which is characterized by the coexistence of the singlet pairing order
parameter $\Delta_{i}\equiv\langle c_{i\downarrow}c_{i\uparrow}\rangle$
and the CDW order $\rho_{\mathbf{Q}}$. As we can see from Fig.~\ref{Flo:pd_12x12},
at fixed on-site interaction strength $U$, the filling region in
which the system remains supersolid increases with the strength of
the dipolar interaction. This can be understood by the following simple
argument. We notice that near the SS-HSF boundary the dominant contribution
to the inter-site interaction energy between two nearest neighbor
sites $i,\, j$ is positive and approximately proportional to $c_{d}\rho_{i}\rho_{j}\sim c_{d}(\rho^{2}-(\rho_{i}-\rho_{j})^{2}/4)$,
therefore a larger $c_{d}$ will make the system more likely to favor
a modulated density distribution since it can lower the intersite
interaction energy in this way. On the other hand, at fixed dipolar
interaction strength $c_{d}$, we can see from Fig.~\ref{Flo:pd_12x12}
that a larger on-site attraction strength $|U|$ corresponds to a
larger supersolid filling factor region. This can be understood by
a similar argument as above. We notice that near the SS-HSF boundary
the total on-site interaction energy of two neighboring sites is approximately
proportional to $-|U|(\rho_{i}^{2}+\rho_{j}^{2})\sim-|U|(\rho^{2}+(\rho_{i}-\rho_{j})^{2}/4)$,
one can thus easily see that for large $|U|$ the system will favor
a large density imbalance between neighboring sites which results
in a modulated density distribution. 

Concerning the proposed experimental setup with $^{40}$K$^{87}$Rb
molecules in an optical lattice, according to the discussions above,
the supersolid phase will be observed more easily in an interaction
region with relatively large dipolar interaction and onsite attraction,
since this will stabilize the supersolid in a large filling factor
region, e.g. in the range $\rho\in(1.0,1.4)$ for interaction strengths
$c_{d}=0.1t$ and $U=-8t$.

On the right-hand side of the SS-HSF boundary, the system is in a
homogeneous singlet superfluid phase which is characterized by a uniform
distribution of both the density $\rho_{i}$ and the singlet paring
order parameter $\Delta_{i}$. 

We note that in the classical limit of zero hopping, the long-range
interacting model exhibits a devil's staircase of various solid phases
\cite{classic_devil_stair_case}. Therefore, for much larger $c_{d}$,
which far exceeds the region investigated in this work, one would
expect to find other types of incompressible CDW phases rather than
CB solid and possibly new SS phases similar to those found for dipolar
Bose gases in optical lattices \cite{dip_bose_square,dip_bose_tri}.
A detailed investigation of this {}``devil's staircase'' in the
quantum case is beyond the scope of this current work.

\subsubsection{Density and singlet pairing order parameter distribution}

\begin{figure}
\includegraphics[clip,width=3.5in]{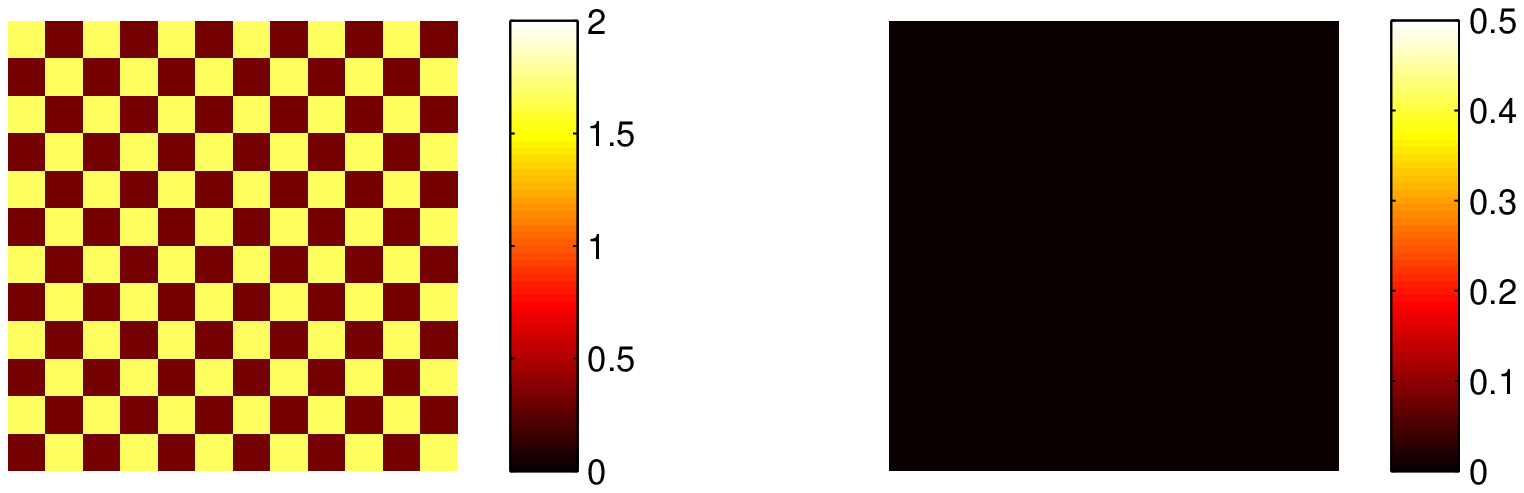}

\includegraphics[clip,width=3.5in]{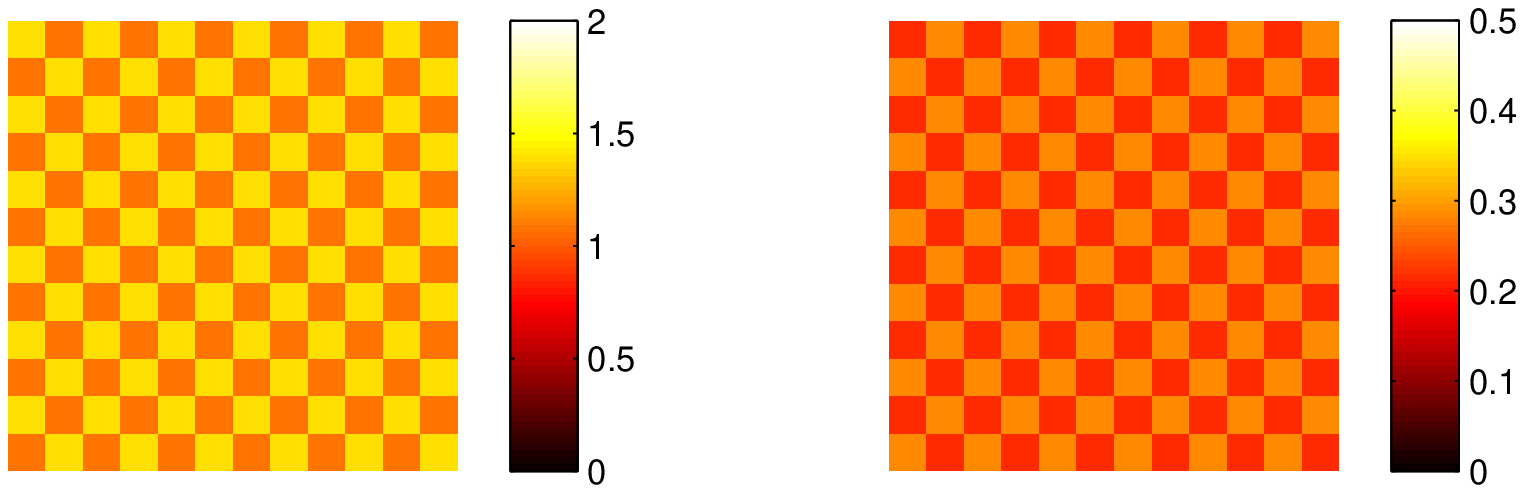}

\includegraphics[clip,width=3.5in]{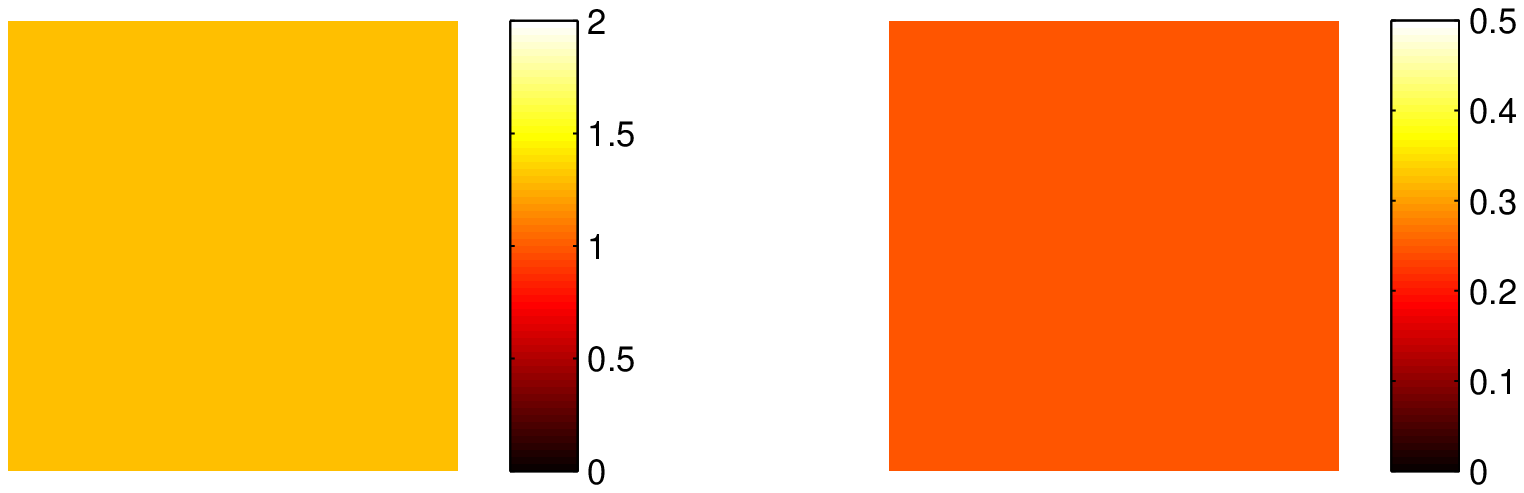}

\caption{(Color online). Zero temperature density $\rho_{i}$ (left panel)
and singlet pairing order parameter $\Delta_{i}$ (right panel) distribution
of the system ($U=-4.0t,\, c_{d}=0.1t$). In descending order, the
filling factor has the values $\rho=1.0,\,1.28,$ and $1.46$ respectively,
corresponding to the checkerboard solid, supersolid, and homogeneous
singlet superfluid. }

\label{Flo:12x12_nf_sc_distribution}
\end{figure}

Recently, there has been considerable progress in single-site addressability
in optical lattices using electron and optical microscopy which allows
for a direct, \emph{in situ, }experimental observation of particle
positions and density-density correlations of the system \cite{Gericke_SAA,Greiner_SAA1,Greiner_SAA2,Bloch_single_atom_address}.
In Fig.~\ref{Flo:12x12_nf_sc_distribution} we show three sets of
snapshots of the density and pairing order parameter distributions
of the system for the different phases discussed previously. We observe
that the system develops the $(\pi,\pi)$ density modulation in both
the CB solid and SS phase. Moreover in the SS phase the amplitude
of the pairing order parameter $\Delta_{i}$ also has a similar spatial
checkerboard modulation. In the HSF phase both the density and the
pairing order parameter are constant over the lattice. We remark here
that although the distribution of density and pairing order parameter
is obtained in the absence of the external trapping potential, the
typical feature of these phases can nevertheless be observed at the
center of a shallow harmonic trap where a local density approximation
is valid.

\subsection{Finite temperature}

In this subsection we investigate finite temperature effects and calculate
the critical temperature of the supersolid which could be useful as
a guide for future experiments. 

\begin{figure}
\includegraphics[clip,width=3in]{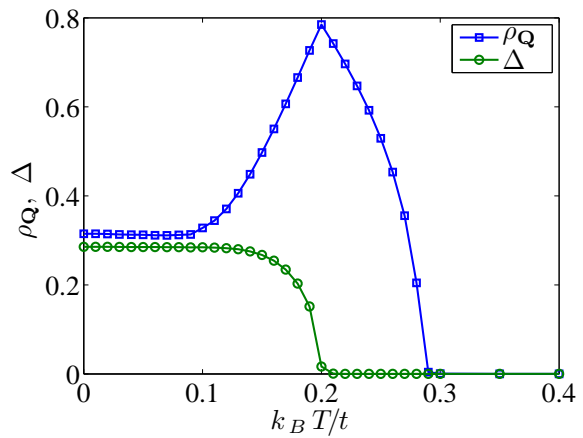}

\caption{(Color online). Melting of the supersolid with increasing temperature
($U=-4.0t$, $c_{d}=0.1t$, $\rho=1.2\,$). Squares and circles correspond
to the temperature dependence of the CDW order $\rho_{\mathbf{Q}}$
and the average value of the pairing order parameters $\Delta$, which
is used to characterize the superfluidity of the system, respectively.}

\label{Flo:SS_melt_U_-4d_nf_1d21}
\end{figure}

First, we study the melting process of the supersolid to a normal
phase with increasing $T$ at fixed filling. Fig.~\ref{Flo:SS_melt_U_-4d_nf_1d21}
shows the temperature dependence of the CDW order parameter $\rho_{\mathbf{Q}}$
and the average value of singlet pairing order parameter $\Delta\equiv\sum_{j}\Delta_{j}/N$,
which is used to characterize the superfluidity of the system, for
$U=-4.0t$, $c_{d}=0.1t$ and filling factor $\rho=1.2\,$. The SS
melts into a normal phase via two successive transitions. First, it
melts into a phase with zero paring order parameter $\Delta_{i}$
but finite $\rho_{\mathbf{Q}}$ at the temperature $T\simeq0.2t/k_{B}$,
where $k_{B}$ is the Boltzmann constant. We will denote this phase
CDW in the following discussion. One interesting behavior to be noted
in this step of the melting process is that as $T$ increases the
CDW order parameter keeps growing to a maximum until the paring order
$\Delta$ decreases to zero, which is quite different from the melting
processes of bosonic supersolids investigated in Ref. \cite{dip_bose_square,dip_bose_tri}.
The physical reason for this behavior is the competition between pairing
and CDW order in the system. Similar phenomena have already been observed
in previous studies of condensed matter systems, see e.g. Ref.~\cite{CDW_BCS_compete}.
To clarify this point, in the vicinity of the phase transition where
both order parameters are small, we calculate the Ginzburg-Landau
(GL) free energy of the system up to the 4th order: \begin{eqnarray}
F & = & A_{\Delta}|\Delta|^{2}+A_{\rho}|\rho_{\mathbf{Q}}|^{2}+B_{\Delta}|\Delta|^{4}+B_{\rho}|\rho_{\mathbf{Q}}|^{4}\nonumber \\
 &  & +C_{\Delta\rho}|\Delta|^{2}|\rho_{\mathbf{Q}}|^{2}.\label{eq:GL_F}\end{eqnarray}
Here $\{A_{\Delta},B_{\Delta},A_{\rho},B_{\rho},C_{\Delta\rho}\}$
are the GL coefficients where a positive $C_{\Delta\rho}$ indicates
competition between the two types of order. A detailed calculation
(see appendix A) shows that $C_{\Delta\rho}$ is indeed positive,
thus justifying the physical picture given above. Upon further increase
of the temperature $T$, the CDW order parameter $\rho_{\mathbf{Q}}$
decreases to zero at $T\simeq0.29t/k_{B}$. 

\begin{figure}
\includegraphics[clip,width=3in]{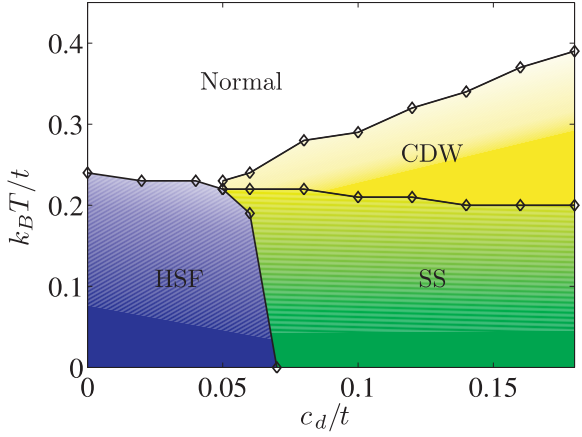}

\caption{(Color online). Finite temperature phase diagram for $U=-4.0t$ and
filling $\rho=1.2$ obtained from calculations on a $12\times12$
square lattice. See the text for details.}

\label{Flo:Finite_temperature_pd}
\end{figure}

Moreover, we map out a phase diagram of the system with respect to
temperature $T$ and the dipolar interaction strength $c_{d}$ for
onsite attraction $U=-4.0t$ at fixed filling $\rho=1.2\,$, see Fig.~\ref{Flo:Finite_temperature_pd}.
When we lower the system temperature, for weak dipolar interaction
($c_{d}<0.05t$) we observe a transition from the normal to the HSF
phase. While in a very narrow region $0.05t<c_{d}<0.07t$, we observe
first a transition between a normal phase (white region) and a CDW
phase (yellow region), then a transition into the SS phase (green
region), and finally a transition into the HSF (blue region) which
is due to the competition between CDW order and pairing as discussed
above. At larger dipolar interaction strength ($c_{d}>0.07t$), we
observe first a transition between the normal phase and the CDW phase,
then a transition into the supersolid. As can be seen from Fig.~\ref{Flo:Finite_temperature_pd},
in order to observe the supersolid for these parameters, $c_{d}$
should be larger than $0.05t$, with a maximum critical temperature
of about $0.22t/k_{B}$ which is approximately $0.1T_{F}\,.$ In comparison
to the lowest temperatures reached in current experiments with polar
molecules in a harmonic trap, which are of order $T_{F}$ \cite{mol_dip_coll},
this means that the temperature still has to be lowered by one order
of magnitude in order to observe the supersolid. But given major theoretical
and experimental efforts in lowering the temperature of cold atoms
and molecules in optical lattices \cite{Cooling_1,Cooling_2,Cooling_3,Cooling_exp1,Cooling_exp2},
we expect this supersolid phase will be accessible experimentally
in the near future.

\section{Conclusion}

In conclusion, we have shown that when fermionic polar molecules are
loaded into a 2D square lattice, this system will exhibit a supersolid
phase which can be observed in an experimental setup based on $^{40}$K$^{87}$Rb
molecules provided the lowest temperature of current experiments can
be lowered by one order of magnitude.
\begin{acknowledgments}
We acknowledge useful discussions with E. Altman, I. Bloch and E.
Demler. L. He thanks I. Titvinidze and A. Privitera for many helpful
discussions and technical support at the beginning of this work. This
work was supported by the German Science Foundation DFG via the DIP
project BL 574/10-1. 
\end{acknowledgments}
\appendix

\section{Calculation of GL coefficients }

In this appendix, we give the detailed calculation of the GL coefficients
from the microscopic Hamiltonian (\ref{eq:Hami}) in the Bloch representation,
i.e., \begin{eqnarray}
H & = & \sum_{\mathbf{k}\sigma}(\varepsilon(\mathbf{k})-\mu)c_{\mathbf{k}\sigma}^{\dagger}c_{\mathbf{k}\sigma}\label{eq:Hami_Bloch}\\
 &  & +\frac{U}{N}\sum_{\mathbf{k}_{1}\mathbf{k}_{2}\mathbf{q}}c_{\mathbf{k}_{1}+\mathbf{q}\uparrow}^{\dagger}c_{\mathbf{k}_{2}-\mathbf{q}\downarrow}^{\dagger}c_{\mathbf{k}_{2}\downarrow}c_{\mathbf{k}_{1}\uparrow}+\frac{1}{2}\sum_{\mathbf{q}}\frac{V(\mathbf{q})}{N}\hat{\rho}_{\mathbf{q}}\hat{\rho}_{-\mathbf{q}}\,,\nonumber \end{eqnarray}
where \begin{eqnarray}
\hat{\rho}_{\mathbf{q}} & = & \sum_{\mathbf{k}\sigma}c_{\mathbf{k}+\mathbf{q},\sigma}^{\dagger}c_{\mathbf{k},\sigma}\;,\\
V(\mathbf{q}) & = & \sum_{|\mathbf{R}_{j}|\neq0}e^{-i\frac{2\pi}{N}\mathbf{q}\cdot\mathbf{R}_{j}}\frac{c_{d}}{|\mathbf{R}_{j}|^{3}}\end{eqnarray}
and $\varepsilon(\mathbf{k})$ is Fourier transformation of the hopping
matrix. The partition function of the system can be written in the
path-integral form of $\mathcal{Z}=\int\mathcal{D}(\bar{c}_{k\sigma},c_{k\sigma})e^{-S[\bar{c}_{k\sigma},c_{k\sigma}]}$
where the action of the system is \begin{eqnarray}
 &  & S[\bar{c}_{k\sigma},c_{k\sigma}]\nonumber \\
 & = & \sum_{k}\sum_{\sigma}(-i\omega_{n}+\varepsilon(\mathbf{k})-\mu)\bar{c}_{k\sigma}c_{k\sigma}\nonumber \\
 &  & +\frac{U}{\beta N}\sum_{k_{1}k_{2}q}\bar{c}_{k_{1}+q\uparrow}\bar{c}_{k_{2}-q\downarrow}c_{k_{2}\downarrow}c_{k_{1}\uparrow}\\
 &  & +\frac{1}{2}\frac{1}{\beta N}\sum_{q}V(\mathbf{q})\sum_{kk'\sigma\sigma'}\bar{c}_{k+q,\sigma}c_{k,\sigma}\bar{c}_{k'-q,\sigma'}c_{k',\sigma'}\nonumber \end{eqnarray}
where $\omega_{n}=(2n+1)\pi/\beta$ is fermionic Matsubara frequency
with $\beta=1/(k_{B}T)$ being the inverse temperature and $q,k$
are three-momentum comprising a vectorial momentum in 2-dimensional
space and a fermionic Matsubara frequency. Now by performing two Hubbard-Stratonovich
transformations\begin{eqnarray}
 &  & \exp[\frac{|U|}{\beta N}\sum_{k_{1}k_{2}q}\bar{c}_{k_{1}+q\uparrow}\bar{c}_{k_{2}-q\downarrow}c_{k_{2}\downarrow}c_{k_{1}\uparrow}]\nonumber \\
 & = & \int\mathcal{D}(\chi,\bar{\chi})\exp\left\{ -\sum_{q}\frac{\beta N}{|U|}|\chi_{q}|^{2}\right.\\
 &  & \left.+\bar{\chi}_{q}\left(\sum_{k}c_{k\downarrow}c_{q-k\uparrow}\right)+\left(\sum_{k}\bar{c}_{q-k\uparrow}\bar{c}_{k\downarrow}\right)\chi_{q}\right\} \,,\nonumber \end{eqnarray}
\begin{eqnarray}
 &  & \exp\left[-\sum_{q}\frac{1}{2}\frac{V(\mathbf{q})}{\beta N}\sum_{kk'\sigma\sigma'}\bar{c}_{k+q,\sigma}c_{k,\sigma}\bar{c}_{k'-q,\sigma'}c_{k',\sigma'}\right]\nonumber \\
 & = & \int\mathcal{D}\phi\exp[-\sum_{q}\frac{\beta N}{2}\phi_{q}V^{-1}(\mathbf{q})\phi_{-q}\nonumber \\
 &  & -i\sum_{q}\phi_{q}\sum_{k\sigma}\bar{c}_{k-q,\sigma}c_{k,\sigma}]\,,\end{eqnarray}
we introduce two auxiliary fields $\chi_{q}$ and $\phi_{q}$ which
separately relate to the singlet pairing order and charge density
wave order in the following way:\begin{eqnarray}
\langle\chi_{q}\rangle & = & \frac{|U|}{\beta N}\langle\sum_{k}c_{k\downarrow}c_{q-k\uparrow}\rangle,\label{eq:pairing_field}\\
\langle\phi_{q}\rangle & = & \frac{2iV(\mathbf{q})}{\beta N}\langle\sum_{k\sigma}\bar{c}_{k+q,\sigma}c_{k,\sigma}\rangle.\label{eq:cdw_field}\end{eqnarray}

After integrating out the fermionic degrees of freedom $\bar{c}_{k\sigma}$
and $c_{k\sigma}$, the partition function becomes $\mathcal{Z}=\int\mathcal{D}(\bar{\chi},\chi,\phi)e^{-S[\bar{\chi},\chi,\phi]}$
where\begin{eqnarray}
S[\bar{\chi},\chi,\phi] & = & \sum_{q}\left(\frac{\beta N}{|U|}|\chi_{q}|^{2}+\frac{\beta N}{2}\phi_{q}V^{-1}(\mathbf{q})\phi_{-q}\right)\nonumber \\
 &  & -\mathrm{tr}\ln\hat{\mathcal{G}}^{-1}\,.\end{eqnarray}
The matrix $\hat{\mathcal{G}}^{-1}$ have the following structure,
\begin{equation}
\hat{\mathcal{G}}^{-1}=\left(\begin{array}{cc}
\mathcal{G}_{0}^{p-1}+\Phi & \Lambda\\
\bar{\Lambda} & -\mathcal{G}_{0}^{p-1}+\tilde{\Phi}\end{array}\right)\end{equation}
where matrices $\mathcal{G}_{0}^{p-1}$, $\Lambda$, $\bar{\Lambda}$,
$\Phi$, and $\tilde{\Phi}$ are given by $\left(\mathcal{G}_{0}^{p-1}\right)_{kk'}=\left(i\omega_{n}-\left(\varepsilon(\mathbf{k})-\mu\right)\right)\delta_{kk'}$,
$\Lambda_{kk'}=\chi_{k+k'}$, $\bar{\Lambda}_{kk'}=\bar{\chi}_{k+k'}$,
$\Phi_{kk'}=-i(\phi_{-k+k'})$, and $\tilde{\Phi}_{kk'}=i(\phi_{k-k'})$. 

Before presenting a further analysis of the effective action $S[\bar{\chi},\chi,\phi]$
in terms of the specific order parameter modes which we are interested
in, one thing to be noted is that the zero mode of $\phi_{q}$, i.e.
$\phi_{(\mathbf{0},0)}$, corresponding to the particle density of
the system, always has a non-vanishing contribution to the effective
action $S[\bar{\chi},\chi,\phi]$ independently of the system's parameters
(interaction strength, temperature, etc). But on the other hand, we
also note that the effect of $\phi_{(\mathbf{0},0)}$ is just to renormalize
the chemical potential $\mu$. Thus in the following analysis we simply
neglect $\phi_{(\mathbf{0},0)}$ in $S[\bar{\chi},\chi,\phi]$ by
replacing the bare chemical potential $\mu$ with a renormalized one
$\bar{\mu}$ which can be determined from the particle density of
the system.

To simplify the notation, we further define\begin{equation}
\hat{\mathcal{G}}_{0}^{-1}\equiv\left(\begin{array}{cc}
\mathcal{G}_{0}^{p-1} & 0\\
0 & -\mathcal{G}_{0}^{p-1}\end{array}\right),\end{equation}
\begin{equation}
\hat{\Phi}\equiv\left(\begin{array}{cc}
\Phi & 0\\
0 & \tilde{\Phi}\end{array}\right),\hat{\Lambda}\equiv\left(\begin{array}{cc}
0 & \Lambda\\
\bar{\Lambda} & 0\end{array}\right).\end{equation}
 We note that \begin{equation}
\mathrm{tr}\ln\hat{\mathcal{G}}^{-1}=\mathrm{tr}\ln\hat{\mathcal{G}}_{0}^{-1}+\mathrm{tr}\ln[1+\hat{\mathcal{G}}_{0}\hat{\Lambda}+\hat{\mathcal{G}}_{0}\hat{\Phi}],\label{eq:trace_ln}\end{equation}
where the first term just contributes a trivial constant. In the vicinity
of the phase transition when both the order parameter fields $\rho_{q}$
and $\Delta_{q}$ are small, we can expand the second term in (\ref{eq:trace_ln})
in terms of $\hat{\mathcal{G}}_{0}\hat{\Lambda}$ and $\hat{\mathcal{G}}_{0}\hat{\Phi}$,
i.e., \begin{equation}
\mathrm{tr}\ln[1+\hat{\mathcal{G}}_{0}\hat{\Lambda}+\hat{\mathcal{G}}_{0}\hat{\Phi}]=\sum_{n=1}\frac{(-1)^{n+1}}{n}\mathrm{tr}(\hat{\mathcal{G}}_{0}\hat{\Lambda}+\hat{\mathcal{G}}_{0}\hat{\Phi})^{n}.\label{eq:tr_ln_expansion}\end{equation}
Odd order terms of either $\hat{\Lambda}$ or $\hat{\phi}$ in the
above expansion vanish since the effective action $S[\bar{\chi},\chi,\phi]$
preserves the symmetry of the unordered phase. Since we are concerned
with finite temperature phase transitions, we neglect the quantum
fluctuations by treating $\chi_{q}$ and $\phi_{q}$ to be independent
of Matsubara frequency, i.e., we focus on the zero Matsubara frequency
component of $\chi_{q}$ and $\phi_{q}$. Furthermore we assume that
$\chi_{(\mathbf{0},0)}$, $\phi_{(\mathbf{Q},0)}$ and their conjugate
dominate the effective interaction $S[\bar{\chi},\chi,\phi]$, where
$\mathbf{Q}=(\pi,\pi)$. From the above expansion formula (\ref{eq:tr_ln_expansion}),
the Ginzburg-Landau coefficients $\{A_{\Delta},B_{\Delta},A_{\rho},B_{\rho},C_{\Delta\rho}\}$
can be calculated explicitly. 

In the following, we calculate the GL coefficient $C_{\Delta\rho}$
which we are most interested in. Expanding $S[\bar{\chi},\chi,\phi]$
to the 4th order in $\chi_{(\mathbf{0},0)}$ and $\phi_{(\mathbf{Q},0)}$
, the cross term between $\chi_{(\mathbf{0},0)}$ and $\phi_{(\mathbf{Q},0)}$
is given by \begin{eqnarray}
 &  & \frac{1}{4}\left\{ 2\mathrm{tr}\left[\left(\hat{\mathcal{G}}_{0}\hat{\Lambda}\hat{\mathcal{G}}_{0}\hat{\Phi}\right)^{2}\right]+4\mathrm{tr}\left[\left(\hat{\mathcal{G}}_{0}\hat{\Lambda}\right)^{2}\left(\hat{\mathcal{G}}_{0}\hat{\Phi}\right)^{2}\right]\right\} \nonumber \\
 & = & \sum_{\mathbf{k},i\omega_{n}}\left[\frac{4}{(\xi^{2}(\mathbf{k})+\omega_{n}^{2})(i\omega_{n}-\xi(\mathbf{k}))(i\omega_{n}-\xi(\mathbf{k+Q}))}\right.\nonumber \\
 &  & \left.+\frac{2}{(\xi^{2}(\mathbf{k})+\omega_{n}^{2})(\xi^{2}(\mathbf{k}+\mathbf{Q})+\omega_{n}^{2})}\right]|\chi_{(\mathbf{0},0)}|^{2}|\phi_{(\mathbf{Q},0)}|^{2}\nonumber \\
 & = & C_{\chi\phi}\,|\chi_{(\mathbf{0},0)}|^{2}|\phi_{(\mathbf{Q},0)}|^{2}\end{eqnarray}
where $\xi(\mathbf{k})=\varepsilon(\mathbf{k})-\bar{\mu}$. It accounts
for the competition between singlet pairing and CDW order. From (\ref{eq:pairing_field})
and (\ref{eq:cdw_field}) we easily obtain $\langle\chi_{(\mathbf{0},0)}\rangle=|U|\Delta$
and $\langle\phi_{q}\rangle=2iV(\mathbf{q})\rho_{\mathbf{Q}}$ which
gives $C_{\Delta\rho}=C_{\chi\phi}4V^{2}(\mathbf{Q})U^{2}$ after
identifying the effective action $S[\bar{\chi},\chi,\phi]$ with the
Ginzburg-Landau free energy. 

\begin{figure}
\includegraphics[clip,width=3in]{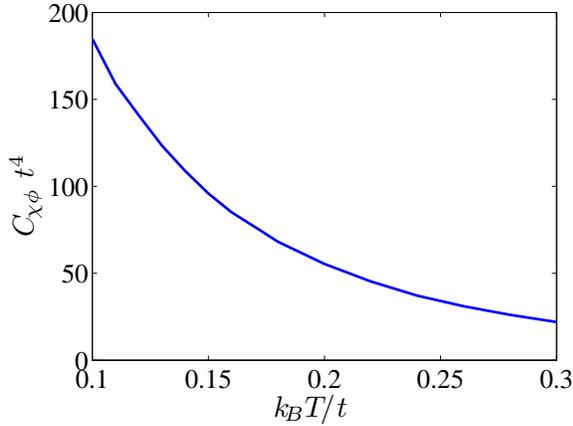}\caption{Temperature dependence of $C_{\chi\phi}$ at filling $\rho=1.2$ .}

\label{Flo:GL4_vs_temperatur_rho_0d6}

\end{figure}

Now we calculate $C_{\chi\phi}$. The summation over Matsubara frequency
can be performed explicitly:\begin{widetext}

\begin{eqnarray}
 &  & \sum_{i\omega_{n}}\left[\frac{4}{(\xi^{2}(\mathbf{k})+\omega_{n}^{2})(i\omega_{n}-\xi(\mathbf{k}))(i\omega_{n}-\xi(\mathbf{k+Q}))}+\frac{2}{(\xi^{2}(\mathbf{k})+\omega_{n}^{2})(\xi^{2}(\mathbf{k}+\mathbf{Q})+\omega_{n}^{2})}\right]\nonumber \\
 & = & (\frac{-\beta}{2\pi i})\ointctrclockwise dz\frac{1}{e^{\beta z}+1}\left[\frac{4}{(z-\xi(\mathbf{k}))^{2}(z+\xi(\mathbf{k}))(z-\xi(\mathbf{k}+\mathbf{Q}))}+\frac{2}{(z^{2}-\xi^{2}(\mathbf{k}))(z^{2}-\xi^{2}(\mathbf{k}+\mathbf{Q}))}\right]\nonumber \\
 & = & \beta\left[\frac{1}{\xi(\mathbf{k})+\xi(\mathbf{k}+\mathbf{Q})}\left(\frac{4}{(1+e^{\beta\xi(\mathbf{k}+\mathbf{Q})})(\xi(\mathbf{k}+\mathbf{Q})-\xi(\mathbf{k}))^{2}}-\frac{e^{\beta\xi(\mathbf{k})}}{(1+e^{\beta\xi(\mathbf{k})})\xi^{2}(\mathbf{k})}\right)\right.\nonumber \\
 &  & +\frac{\xi(\mathbf{k}+\mathbf{Q})-3\xi(\mathbf{k})+e^{\beta\xi(\mathbf{k})}(\xi(\mathbf{k}+\mathbf{Q})+\xi(\mathbf{k})(-3+2\beta(\xi(\mathbf{k}+\mathbf{Q})-\xi(\mathbf{k}))))}{(1+e^{\beta\xi(\mathbf{k})})^{2}\xi^{2}(\mathbf{k})(\xi(\mathbf{k}+\mathbf{Q})-\xi(\mathbf{k}))^{2}}\nonumber \\
 &  & \left.+\frac{1}{\xi^{2}(\mathbf{k})-\xi^{2}(\mathbf{k}+\mathbf{Q})}\left(\frac{\tanh\frac{\xi(\mathbf{k}+\mathbf{Q})}{2}}{\xi(\mathbf{k}+\mathbf{Q})}-\frac{\tanh\frac{\xi(\mathbf{k})}{2}}{\xi(\mathbf{k})}\right)\right].\end{eqnarray}
\end{widetext} The remaining summation over momentum $\mathbf{k}$
can be easily replaced by an integral over the density of states $\rho_{\mathrm{DOS}}(\varepsilon)\equiv\sum_{\mathbf{k}}\delta(\varepsilon-\varepsilon(\mathbf{k}))$
and evaluated numerically. As we can see from Fig.~\ref{Flo:GL4_vs_temperatur_rho_0d6},
which shows the temperature dependence of $C_{\chi\phi}$ when $0.1<k_{B}T/t<0.3$
at filling $\rho=1.2$, $C_{\chi\phi}$ is positive in this temperature
region. Therefore, since the GL coefficient $C_{\Delta\rho}$ has
the same sign as $C_{\chi\phi}$ , we now clearly see that singlet
pairing and CDW order indeed compete with each other in the parameter
region under investigation.

\begin{thebibliography}{60}
\bibitem{BH_The} D. Jaksch, C. Bruder, J. I. Cirac, C. W. Gardiner,
and P. Zoller, Phys. Rev. Lett. \textbf{81}, 3108 (1998).

\bibitem{BH_Exp}M. Greiner, O. Mandel, T. Esslinger, T.W. H\"{a}nsch,
and I. Bloch, Nature \textbf{415}, 39 (2002).

\bibitem{Lattice_review}I. Bloch, J. Dalibard, and W. Zwerger, Rev.
Mod. Phys. \textbf{80}, 885 (2008).

\bibitem{Fermi_Hubbard_The}W. Hofstetter, J. I. Cirac, P. Zoller,
E. Demler and M. D. Lukin, Phys. Rev. Lett. \textbf{89}, 220407 (2002). 

\bibitem{Fermi_Hubbard_Exp}R. J\"{o}rdens, N. Strohmaier, K. G\"{u}nter,
H. Moritz, and T. Esslinger, Nature \textbf{455}, 204 (2008). 

\bibitem{dip_the_1}S. Yi and L. You, Phys. Rev. A \textbf{61}, 041604
(2000).

\bibitem{dip_the_2}S. Yi and L. You, Phys. Rev. A \textbf{63}, 053607
(2001).

\bibitem{dip_the_3}K. Goral, K. Rzazewski, and T. Pfau, Phys. Rev.
A \textbf{61}, 051601 (2000).

\bibitem{dip_the_4}L. Santos, G. V. Shlyapnikov, P. Zoller, and M.
Lewenstein, Phys. Rev. Lett. \textbf{85}, 1791 (2000).

\bibitem{dip_the_5}S. Ronen, D. C. E. Bortolotti, and J. L. Bohn,
Phys. Rev. Lett. \textbf{98}, 030406 (2007).

\bibitem{dip_exp1}A. Griesmaier, J. Werner, S. Hensler, J. Stuhler,
and T. Pfau, Phys. Rev. Lett. \textbf{94}, 160401 (2005).

\bibitem{dip_exp2}J. Stuhler, A. Griesmaier, T. Koch, M. Fattori,
T. Pfau, S. Giovanazzi, P. Pedri, and L. Santos, Phys. Rev. Lett.
\textbf{95}, 150406 (2005).

\bibitem{dip_exp_3}A. Griesmaier, J. Stuhler, T. Koch, M. Fattori,
T. Pfau, and S. Giovanazzi, Phys. Rev. Lett. \textbf{97}, 250402 (2006).

\bibitem{dip_exp_4}T. Lahaye, J. Metz, B. Fr\"{o}hlich, T. Koch,
M. Meister, A. Griesmaier, T. Pfau, H. Saito, Y. Kawaguchi, and M.
Ueda, Phys. Rev. Lett. \textbf{101}, 080401 (2008).

\bibitem{dip_bose_lat_mf_1} K. Goral, L. Santos, and M. Lewenstein,
Phys. Rev. Lett. \textbf{88}, 170406 (2002).

\bibitem{dip_bose_lat_mf_2} S. Yi, T. Li, and C. P. Sun, Phys. Rev.
Lett. \textbf{98}, 260405 (2007).

\bibitem{dip_bose_square} B. Capogrosso-Sansone, C. Trefzger, M.
Lewenstein, P. Zoller, and G. Pupillo, Phys. Rev. Lett. \textbf{104},
125301 (2010).

\bibitem{dip_bose_tri}L. Pollet, J. D. Picon, H. P. B\"{u}chler, and
M. Troyer, Phys. Rev. Lett. \textbf{104}, 125302 (2010).

\bibitem{dip_fermi_1}K. G\'{o}ral, B.-G. Englert, and K. Rza\.{z}ewski,
Phys. Rev. A \textbf{63}, 033606 (2001).

\bibitem{dip_fermi_2}M. A. Baranov, M. S. Marenko, Val. S. Rychkov,
and G. V. Shlyapnikov, Phys. Rev. A \textbf{66}, 013606 (2002).

\bibitem{dip_fermi_3}M. A. Baranov, \L{}. Dobrek, and M. Lewenstein,
Phys. Rev. Lett. \textbf{92}, 250403 (2004).

\bibitem{dip_fermi_4}M. A. Baranov, K. Osterloh, and M. Lewenstein,
Phys. Rev. Lett. \textbf{94}, 070404 (2005).

\bibitem{dip_fermi_5}K. Osterloh, N. Barber\'{a}n, and M. Lewenstein,
Phys. Rev. Lett. \textbf{99}, 160403 (2007).

\bibitem{PM_Exp} K.-K. Ni, S. Ospelkaus, M. H. G. de Miranda, A.
Pe'er, B. Neyenhuis, J. J. Zirbel, S. Kotochigova, P. S. Julienne,
D. S. Jin, and J. Ye, Science \textbf{322}, 231 (2008).

\bibitem{dip_fer_molecule_1}L. He, J. N. Zhang, Y. Zhang, and S.
Yi, Phys. Rev. A \textbf{77}, 031605(R) (2008). 

\bibitem{dip_fer_molecule_2}T. Miyakawa, T. Sogo, and H. Pu, Phys.
Rev. A \textbf{77}, 061603(R) (2008). 

\bibitem{dip_fer_molecule_3}G. M. Bruun and E. Taylor, Phys. Rev.
Lett. \textbf{101}, 245301 (2008). 

\bibitem{dip_fer_molecule_4}T. Sogo, L. He, T. Miyakawa, S. Yi, H.
Lu, and H. Pu, New J. Phys. \textbf{11}, 055017 (2009). 

\bibitem{dip_fer_molecule_5} J.-N. Zhang and S. Yi, Phys. Rev. A
\textbf{80}, 053614 (2009). 

\bibitem{dip_fer_molecule_6} K. Sun, C. -J. Wu, and S. Das Sarma, Phys. Rev.
B \textbf{82}, 075105 (2010).

\bibitem{dip_fer_molecule_7}Y. Yamaguchi, T. Sogo, T. Ito, and T.
Miyakawa, Phys. Rev. A \textbf{82}, 013643 (2010). 

\bibitem{dip_fer_molecule_8}C. Zhao, L. Jiang, X.-X. Liu, W. M. Liu,
X.-B. Zou, and H. Pu, Phys. Rev. A \textbf{81}, 063642 (2010). 

\bibitem{dip_fer_molecule_9}A. R. P. Lima and A. Pelster, Phys. Rev.
A \textbf{81}, 063629 (2010).

\bibitem{dip_fer_molecule_10}Y. Endo, T. Miyakawa, and T. Nikuni,
Phys. Rev. A \textbf{81}, 063624 (2010).

\bibitem{dip_fer_molecule_11}J.-N. Zhang and S. Yi, Phys. Rev. A
\textbf{81}, 033617 (2010). 

\bibitem{dip_fer_molecule_12}A. R. P. Lima and A. Pelster, Phys.
Rev. A \textbf{81}, 021606 (2010). 

\bibitem{dip_fer_molecule_13}C.-K. Chan, C.-J. Wu, W.-C. Lee, and
S. Das Sarma, Phys. Rev. A \textbf{81}, 023602 (2010). 

\bibitem{dip_fer_molecule_14}S. Ronen and J. L. Bohn, Phys. Rev.
A \textbf{81}, 033601 (2010). 

\bibitem{dip_fer_molecule_15}C.-H. Lin, Y.-T. Hsu, H. Lee, and D.-W.
Wang, Phys. Rev. A \textbf{81}, 031601 (2010). 

\bibitem{dip_fer_molecule_16}C.-W. Lin, E. Zhao, and W. V. Liu, Phys.
Rev. B \textbf{81}, 045115 (2010). 

\bibitem{dip_fer_molecule_17}C.-J. Wu and J. E. Hirsch, Phys. Rev.
B \textbf{81}, 020508 (2010).

\bibitem{dip_fer_molecule_18}T. Shi, J.-N. Zhang, C.-P. Sun, and S. Yi, Phys. Rev. A {\bf82}, 033623 
(2010).

\bibitem{dip_fer_molecule_19}K. Mikelsons and J. K. Freericks, Phys. Rev. A {\bf83}, 043609 (2011)

\bibitem{mol_dip_coll}K.-K. Ni, S. Ospelkaus, D. Wang, G. Qu\'{e}m\'{e}ner,
B. Neyenhuis, M. H. G. de Miranda, J. L. Bohn, J. Ye, and D. S. Jin,
Nature \textbf{464}, 1324 (2010). 

\bibitem{R-DMFT_1}M. Snoek, I. Titvinidze, C. Toke, K. Byczuk, and
W. Hofstetter, New J. Phys. \textbf{10}, 093008 (2008).

\bibitem{R-DMFT_2} R. W. Helmes, T. A. Costi, and A. Rosch, Phys.
Rev. Lett. \textbf{100}, 056403 (2008).

\bibitem{DMFT} A. Georges, G. Kotliar, W. Krauth, and M. J. Rozenberg,
Rev. Mod. Phys. \textbf{68}, 13 (1996). 

\bibitem{CDMFT}G. Kotliar, S. Y. Savrasov, G. P\'{a}lsson, and G. Biroli,
Phys. Rev. Lett. \textbf{87}, 186401 (2001). 

\bibitem{LD_HFA}E. M\"{u}ller-Hartmann, Z. Phys. B \textbf{74},\textbf{
}507 (1989).

\bibitem{ED_solver_1}M. Caffarel and W. Krauth, Phys. Rev. Lett.
\textbf{72}, 1545 (1994).

\bibitem{ED_solver_2}Q.-M. Si, M. J. Rozenberg, G. Kotliar, and A.
E. Ruckenstein, Phys. Rev. Lett. \textbf{72}, 2761 (1994).

\bibitem{classic_devil_stair_case}P. Bak and R. Bruinsma, Phys. Rev.
Lett. \textbf{49}, 249 (1982).

\bibitem{Gericke_SAA} T. Gericke, P. W\"{u}rtz, D. Reitz, T. Langen,
and H. Ott, Nature Phys. \textbf{4}, 949 (2008). 

\bibitem{Greiner_SAA1} W. S. Bakr, J. I. Gillen, A. Peng, S. F\"{o}lling,
and M. Greiner, Nature \textbf{462}, 74 (2009). 

\bibitem{Greiner_SAA2}W. S. Bakr, A. Peng, M. E. Tai, R. Ma, J. Simon,
J. I. Gillen, S. F\"{o}lling, L. Pollet, and M. Greiner, Science\textbf{
329}, 5991 (2010). 

\bibitem{Bloch_single_atom_address}J. F. Sherson, C. Weitenberg,
M. Endres, M. Cheneau, I. Bloch, S. Kuhr, Nature \textbf{467}, 68
(2010).

\bibitem{CDW_BCS_compete} S. Robaszkiewicz, R. Micnas, and K. A.
Chao, Phys. Rev. B \textbf{24}, 1579 (1981). 

\bibitem{Cooling_1} M. Popp, J.-J. Garcia-Ripoll, K. G. Vollbrecht,
and J. I. Cirac, Phys, Rev. A \textbf{74}, 013622 (2006). 

\bibitem{Cooling_2} T. Ho, Q. Zhou, Natl. Acad. Sci. U.S.A. \textbf{106},
6916 (2009).

\bibitem{Cooling_3} J. Catani, G. Barontini, G. Lamporesi, F. Rabatti,
G. Thalhammer, F. Minardi, S. Stringari, and M. Inguscio, Phys. Rev.
Lett. \textbf{103}, 140401 (2009).

\bibitem{Cooling_exp1} D. M. Weld, P. Medley, H. Miyake, D. Hucul,
D. E. Pritchard, and W. Ketterle, Phys. Rev. Lett. \textbf{103}, 245301
(2009).

\bibitem{Cooling_exp2}P. Medley, D. M. Weld, H. Miyake, D. E. Pritchard,
and W. Ketterle, arXiv:1006.4674.
\end{thebibliography}
\end{document}